\documentclass[
superscriptaddress,
reprint,
 amsmath,amssymb,
 aps,
 prl,
]{revtex4-2}

\usepackage{graphicx}
\usepackage{dcolumn}
\usepackage{bm}

\begin{document}

\title{Femtosecond Core-Level Charge Transfer}

\author{Simon P. Neville}
\email{Simon.Neville@nrc-cnrc.gc.ca}
\affiliation{National Research Council Canada, 100 Sussex Drive,
  Ottawa, Ontario K1A 0R6, Canada}

\author{Martha Yaghoubi Jouybari}
\affiliation{Department of Chemistry and Biomolecular Sciences,
  University of Ottawa, 10 Marie Curie, Ottawa, Ontario, K1N 6N5,
  Canada}

\author{Michael S. Schuurman}
\affiliation{National Research Council Canada, 100 Sussex Drive,
  Ottawa, Ontario K1A 0R6, Canada}
\affiliation{Department of Chemistry and Biomolecular Sciences,
  University of Ottawa, 10 Marie Curie, Ottawa, Ontario, K1N 6N5,
  Canada}

\date{\today}

\begin{abstract}
Charge transfer is a fundamental phenomenon in biology and chemistry, and involves the movement of charge through a system driven by nuclear dynamics. Because of the involvement of nuclear motion, it is generally assumed that charge transfer will occur on a time-scale of some few tens-to-hundreds of femtoseconds. Using the example of ethylene excited to its $1s\pi^{*}$ manifold, we demonstrate that ultrafast, few-femtosecond core-level charge transfer may occur following core-excitation, driven by the formation of electronic coherences by non-adiabatic dynamics. Here, transfer of core-electron density from one side of the molecule to the other is driven by a breakdown of the Born-Oppenheimer approximation, and results in core-hole localisation occuring within 5~fs, followed by core-hole delocalisation, all within the Auger decay window. These results serve to demonstrate that ultra-fast core-level charge transfer driven by nuclear dynamics may occur on the same timescale as purely electronic, i.e., charge migration, dynamics following core-excitation.
\end{abstract}

\maketitle
The observation of ultrafast electron dynamics in molecules is a long-standing goal in atomic, molecular and optical physics\cite{worner2017}. Particular attention has been paid to the phenomenon of charge migration following ultrafast ionisation of a molecular system. Here, the creation of a coherent superposition of electronic states leads to the ultrafast movement of the ``hole" in the electron density through the system, and is an entirely electronic phenomenon\cite{cederbaum1999,cederbaum2003}. Charge migration is operative following both valence\cite{kuleff2011,kuleff2015,kuleff2018,lopata2021,kraus2015,dey2022} and core\cite{kuleff2016,vendrell2024,inhester2019} excitation, and is driven by the formation of electronic coherences by a pump laser pulse. However, coupling to the nuclear degrees of freedom can result in a rapid electronic decoherence, leading to a termination of the charge migration dynamics\cite{vacher2015,vacher2017}. Another mechanism for the redistribution of the hole in the electron density following photo-excitation is charge transfer which, by conventional definition, is instead \textit{driven} by nuclear motion, and is thus assumed to occur on a timescale of some few-to-hundreds of femtoseconds (fs). That is, orders of magnitude slower than charge migration. We here demonstrate that, contrary to conventional wisdom, ultrafast (few-fs) charge transfer may, in fact, occur.

We consider X-ray absorption in molecules containing two or more equivalent core atomic orbitals (AOs). This covers broad classes of organic and inorganic molecules of fundamental importance, e.g., the polyenes, acenes, and many transition metal complexes of importance. For $N$ equivalent core AOs, core-excitation will give rise to sets of $N$ near-degenerate core-excited states, each corresponding to excitation to the same final virtual orbital. It is known that strong vibronic coupling exists within such manifolds of nearly-degenerate core-excited states, and that this can result in core-hole localistion in the corresponding vibronic states\cite{gadea1991,koppel1997}. This raises the appealing prospect of broadband X-ray absorption giving rise to core-hole redistribution driven by nuclear motion. That is, core-level charge transfer. We here show that this is indeed that case, and that core-level charge transfer can occur on an ultrafast (few-fs) timescale. Importantly, we demonstrate that these core-hole dynamics occur within the Auger decay window and are therefore, in principle, observable.

\begin{figure*}
    \centering
    \includegraphics[width=1.0\textwidth]{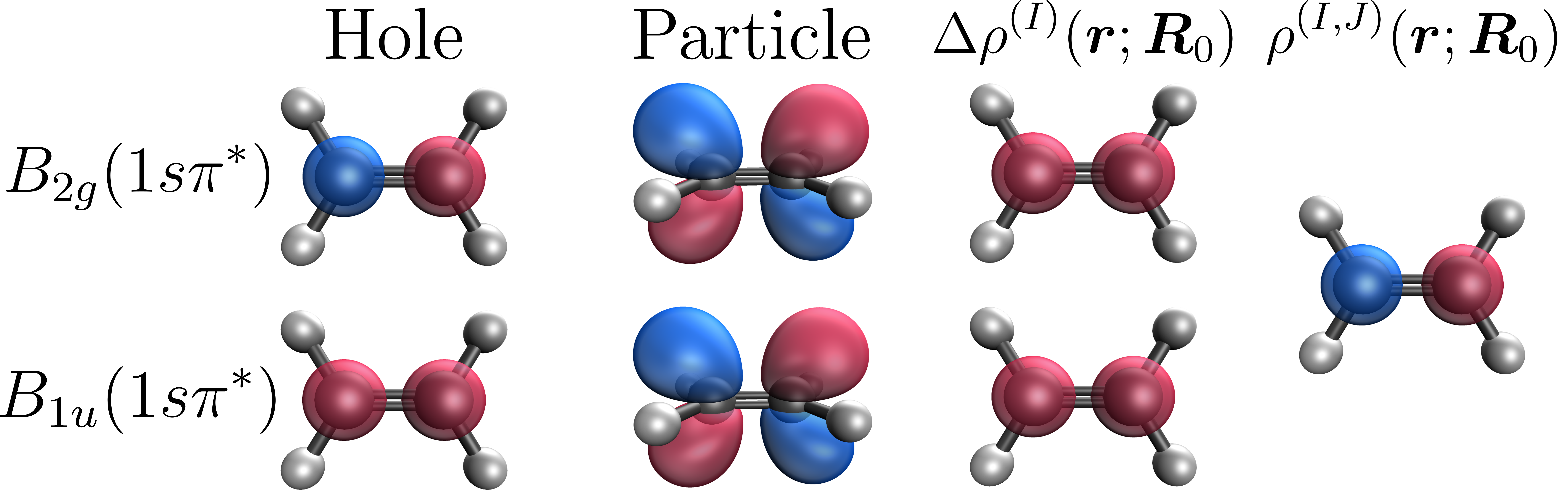}
    \caption{Columns 1 and 2: diabatic hole and particle NTOs computed for the $B_{1u}(1s\pi^{*})$ and $B_{2g}(1s\pi^{*})$ states. Column 3: diabatic one-electron difference densities $\Delta\rho^{(I)}(\boldsymbol{r};\boldsymbol{R}_{0})$ between the $B_{1u}(1s\pi^{*})$ and $B_{2g}(1s\pi^{*})$ states and the ground state. Column 4: diabatic one-electron transition density $\rho^{(I,J)}(\boldsymbol{r}; \boldsymbol{R}_{0})$ between the $B_{1u}(1s\pi^{*})$ and $B_{2g}(1s\pi^{*})$ states. All quantities were calculated at the $D_{2h}$ ground state minimum energy geometry.}
    \label{fig:densities}
\end{figure*}

As a representative example, we consider $1s \rightarrow \pi^{*}$ excitation in ethylene. At the $D_{2h}$ ground state minimum energy geometry, there exit two equivalent C $1s$ AOs. The symmetric and anti-symmetric linear combinations of these gives rise to two near-degenerate C $1s$ molecular orbitals (MOs) of $a_{g}$ and $b_{3u}$ symmetry. Excitation from these core-level orbitals to the $\pi^{*}$ orbital gives rise to the nearly-degenerate $B_{1u}(1s\pi^{*})$ and $B_{2g}(1s\pi^{*})$ states. For reference, the dominant natural transition orbitals (NTOs) for transition to these states are shown in the first two columns of Figure~\ref{fig:densities}. The $B_{1u}(1s\pi^{*})$ state is optically bright, while the $B_{2g}(1s\pi^{*})$ state is optically dark. However, these two states are known to be strongly vibronically coupled by the $b_{3u}$ asymmetric in-plane C-H stretching mode\cite{gadea1991,koppel1997}, leading to vibronic states of mixed $B_{1u}(1s\pi^{*})$ and $B_{2g}(1s\pi^{*})$ character. As we shall demonstrate, excitation to the former using a broadband X-ray pulse leads to the ultrafast transfer of core-electron density from one side of the molecule to the other, resulting in core-hole localisation. This is driven by nuclear motion and is a result of the breakdown of the Born-Oppenheimer approximation and, as such, corresponds to a charge transfer process.

We begin with the definition of some notation. Let $\boldsymbol{r}$ and $\boldsymbol{R}$ denote the vectors of electronic and nuclear degrees of freedom, respectively, and $\Psi(\boldsymbol{r},\boldsymbol{R},t)$ the time-dependent vibronic wave function. Our starting point is the Born-Huang expansion of $\Psi(\boldsymbol{r},\boldsymbol{R},t)$ in terms of a set of electronic wave functions $\{ \psi_{I}(\boldsymbol{r};\boldsymbol{R}) \}$:

\begin{equation}\label{eq:BHE}
    \Psi(\boldsymbol{r},\boldsymbol{R},t) = \sum_{I} \chi_{I}(\boldsymbol{R},t) \psi_{I}(\boldsymbol{r};\boldsymbol{R}).
\end{equation}

Here, and in the following, we shall assume a \textit{diabatic} representation, as the conservation of the characters of the diabatic electronic wave functions with nuclear geometry allows for a simpler interpretation. The parametric dependence of the time-independent electronic wave functions on the nuclear coordinates results in expansion coefficients $\chi_{I}(\boldsymbol{R},t)$ that are both nuclear-coordinate- and time-dependent. These are the nuclear wave functions. For the case considered here, the expansion in Equation~\ref{eq:BHE} shall be limited to the $B_{1u}(1s\pi^{*})$ and $B_{2g}(1s\pi^{*})$ states. The time-evolution of the nuclear wave functions encodes any core-level dynamics that may be initiated by X-ray excitation. The question, then, is how to quantify and, ideally, visualise these dynamics. To this end, we note that the Born-Huang wave function \textit{ansatz} allows us to define a \textit{vibronic} one-electron reduced density (1-VRD)

\begin{equation}\label{eq:1-vrd}
    \rho(\boldsymbol{r}_{1}, \boldsymbol{R}, t) = n_{el} \int d\boldsymbol{r}_{2} \cdots \boldsymbol{r}_{n_{el}} \Psi^{*}(\boldsymbol{r},\boldsymbol{R},t) \Psi(\boldsymbol{r},\boldsymbol{R},t),
\end{equation}

\noindent
where $n_{el}$ is the number of electrons. For the sake of notational brevity, and acknowledging that the electrons are indistinguishable, we shall use $\mathbf{r}$ to denote the coordinates of a single electron instead of $\boldsymbol{r}_{1}$. The 1-VRD contains information about the coupled electron-nuclear dynamics, from which the vibronic core-level dynamics may be extracted.

Inserting, the Born-Huang expansion of the wave function into Equation~\ref{eq:1-vrd}, we obtain the following expression for the 1-VRD:

\begin{equation}\label{eq:1-vrd_2}
    \rho(\mathbf{r},\boldsymbol{R},t) = \sum_{I,J} \chi_{I}^{*}(\boldsymbol{R},t) \chi_{J}(\boldsymbol{R},t) \rho^{(I,J)}(\textbf{r};\boldsymbol{R}),
\end{equation}

\noindent
where the $\rho^{(I,J)}(\mathbf{r};\boldsymbol{R})$ are the standard electronic one-electron reduced (transition) densities:

\begin{equation}
    \rho^{(I,J)}(\textbf{r};\boldsymbol{R}) = n_{el} \int d\boldsymbol{r}_{2} \cdots d\boldsymbol{r}_{n_{el}} \psi_{I}^{*}(\boldsymbol{r};\boldsymbol{R}) \psi_{J}(\boldsymbol{r};\boldsymbol{R}).
\end{equation}

We next introduce the concept of a ``vibronic core-hole"". An intuitive definition is as follows. The difference between 1-VRD $\rho(\mathbf{r},\boldsymbol{R},t)$ and the diabatic ground state one-electron reduced density, $\rho_{0}(\mathbf{r};\boldsymbol{R})$, is first computed, yielding the vibronic one-electron difference density $\Delta \rho(\mathbf{r},\boldsymbol{R},t)$:

\begin{equation}\label{eq:diff_dens}
    \Delta \rho(\mathbf{r},\boldsymbol{R},t) = \rho(\mathbf{r},\boldsymbol{R},t) - \alpha(\boldsymbol{R},t) \rho_{0}(\mathbf{r};\boldsymbol{R}),
\end{equation}

\begin{equation}
    \alpha(\boldsymbol{R},t) = \frac{1}{n_{el}} \int d\mathbf{r} \rho(\mathbf{r},\boldsymbol{R},t),
\end{equation}

\noindent
where the scaling factor $\alpha(\boldsymbol{R},t)$ ensures that $\Delta \rho(\mathbf{r},\boldsymbol{R},t)$ integrates to zero, as should be the case for an electron-number conserving excitation process. The ``core-hole" is then identified with the vibronic core-hole density, $\Delta \rho^{(core)}(\mathbf{r},\boldsymbol{R},t)$, defined by 

\begin{equation}\label{eq:core_hole_dens}
    \Delta \rho^{(core)}(\mathbf{r},\boldsymbol{R},t) = \hat{P}_{1s}(\mathbf{r};\boldsymbol{R}) \Delta \rho(\mathbf{r},\boldsymbol{R},t),
\end{equation}

\noindent
where $\hat{P}_{1s}(\mathbf{r};\boldsymbol{R})$ is the projector onto the subspace spanned by the $1s$ core-level electrons. Inserting the Equation~\ref{eq:1-vrd_2} into Equation~\ref{eq:core_hole_dens}, and introducing the one-electron difference density $\Delta\rho^{(I)}(\mathbf{r};\boldsymbol{R})$,

\begin{equation}
    \Delta\rho^{(I)}(\mathbf{r};\boldsymbol{R}) = \rho^{(I,I)}(\textbf{r};\boldsymbol{R}) - \rho_{0}(\mathbf{r};\boldsymbol{R}).
\end{equation}

\noindent
we obtain

\begin{equation}
    \Delta \rho^{(core)}(\mathbf{r},\boldsymbol{R},t) = \Delta \rho^{(core)}_{pop}(\mathbf{r},\boldsymbol{R},t) + \Delta \rho^{(core)}_{coh}(\mathbf{r},\boldsymbol{R},t),
\end{equation}

\noindent
with

\begin{equation}
    \Delta \rho_{pop}^{(core)}(\mathbf{r},\boldsymbol{R},t) = \hat{P}_{1s}(\mathbf{r};\boldsymbol{R}) \sum_{I} \left| \chi_{I}(\boldsymbol{R},t) \right|^{2} \Delta\rho^{(I)}(\mathbf{r};\boldsymbol{R}),
\end{equation}

\begin{equation}\label{eq:rho_coh}
    \Delta \rho_{coh}^{(core)}(\mathbf{r},\boldsymbol{R},t) = \hat{P}_{1s}(\mathbf{r};\boldsymbol{R}) \sum_{I \ne J} \chi_{I}^{*}(\boldsymbol{R},t) \chi_{J}(\boldsymbol{R},t) \rho^{(I,J)}(\textbf{r};\boldsymbol{R})
\end{equation}

\noindent
That is, the vibronic core-hole density can be decomposed into two contributions: (i) a ``population" term, $\Delta\rho_{pop}^{(core)}(\mathbf{r},\boldsymbol{R},t)$, involving the one-electron difference densities of the core-excited states involved weighted by the corresponding nuclear probability densities, and; (ii) a ``coherence" term, $\Delta\rho_{coh}^{(core)}(\mathbf{r},\boldsymbol{R},t)$, involving the one-electron transition densities between the core-excited states weighted by the products of the two corresponding nuclear wave functions.

At this point, it is instructive to consider the specific forms of the one-electron difference and transition densities for the $B_{1u}(1s\pi^{*})$ and $B_{2g}(1s\pi^{*})$ states of ethylene following projection onto the core-orbital subspace. These are shown in Figure~\ref{fig:densities} as computed at the $D_{2h}$ ground state minimum energy geometry, denoted here as $\boldsymbol{R}_{0}$. The third column shows the one-electron difference densities, $\Delta\rho^{(I)}(\boldsymbol{r};\boldsymbol{R})$, between the two core-excited states and the ground state, while the fourth column shows the one-electron transition density, $\rho^{(I,J)}(\boldsymbol{r};\boldsymbol{R})$, between the two core-excited states. The one-electron difference densities are essentially identical. Because the difference densities contribute to the vibronic core-hole density with positive weights, it is thus clear that the population contribution, $\Delta\rho_{pop}^{(core)}(\textbf{r},\boldsymbol{R},t)$, alone will always correspond to a delocalised core-hole. That is, a core-hole that is symmetrically distributed over the two carbon atoms. However, the transfer of core-electron density from one side of the molecule to the other may occur if the coherence contribution, $\Delta\rho_{coh}^{(core)}(\mathbf{r},\boldsymbol{R},t)$, is non-negligible. This can be seen from the form of the one-electron transition density $\rho^{(I,J)}(\boldsymbol{r};\boldsymbol{R})$, which contains both positive and negative regions. When added to the population contribution this can lead to the depletion of electron density in the region surrounding one of the two equivalent carbon atoms. That is core-level charge transfer may be driven, leading to core-hole localization. Considering Equation~\ref{eq:rho_coh}, whether or not core-hole localization occurs requires and that the nuclear wave functions on the different electronic states exhibit a high degree of overlap. Whether or not this will happen is contingent on the coupled nuclear-electronic dynamics and, in particular, on whether the two corresponding nuclear wave functions overlap significantly during its course.

To simulate the coupled electronic-nuclear dynamics, we adopt a previously demonstrated strategy for reproducing the population dynamics in a reduced vibrational space employing two effective modes: the gradient difference and non-adiabatic coupling directions, denoted $x$ and $y$, respectively\cite{ryabinkin2014}. Due to the fact that (i) the $B_{1u}(1s\pi^{*})$ and $B_{2g}(1s\pi^{*})$ states are quasi-degenerate at the Frank-Condon point, and (ii) the core-excited states have very short (sub-10~fs) lifetimes due to Auger decay pathways, it is reasonable to expect the nuclear dynamics to be effectively confined to this two-dimensional subspace, which would correspond to a conical intersection branching space if the two states were truly degenerate. The validity of this approximation is demonstrated in the Supplementary Information. 

Our model Hamiltonian corresponds to a vibronic coupling Hamiltonian\cite{koppel1984,cederbaum1977}, and reads as follows:

\begin{equation}
    \boldsymbol{H} = \hat{T} \boldsymbol{1}_{2} + \sum_{n=1}^{6} \boldsymbol{W}^{(n)}(x,y).
\end{equation}

\noindent
Here, $\hat{T}$ denotes the nuclear kinetic energy operator, and $\boldsymbol{W}^{(n)}(x,y)$ the $n$th-order contribution to the Taylor expansion of the diabatic potential matrix. That is, our model corresponds to a sixth-order expansion of the diabatic potential in terms of the effective modes $x$ and $y$. The parameters of the model Hamiltonian were computed using the QD-DFT/MRCI(2) level of theory\cite{qd-dftmrci2,dftmrci2} within the core-valence separation (CVS) approximation\cite{cederbaum1980,barth1981,cvs-dftmrci}. The nuclear wave functions $\chi_{I}(x,y,t)$ were represented in terms of a numerically exact expansion in terms of discrete variable representation basis functions, and the time-dependent Schr\"{o}dinger equation solved using the short iterative Lanczos algorithm\cite{park1986}. The full details of these calculations are given in the Supplementary Information. The initial wave packet, chosen as the vertical excitation of the ground vibronic state to the optically bright $B_{1u}(1s\pi^{*})$ state, was selected to emulate broadband excitation with a few-fs X-ray laser pulse. That this is a good approximation of the initial state prepared by such a pulse is verified in the Supplementary Information.

\begin{figure}
    \centering
    \includegraphics[width=0.5\textwidth]{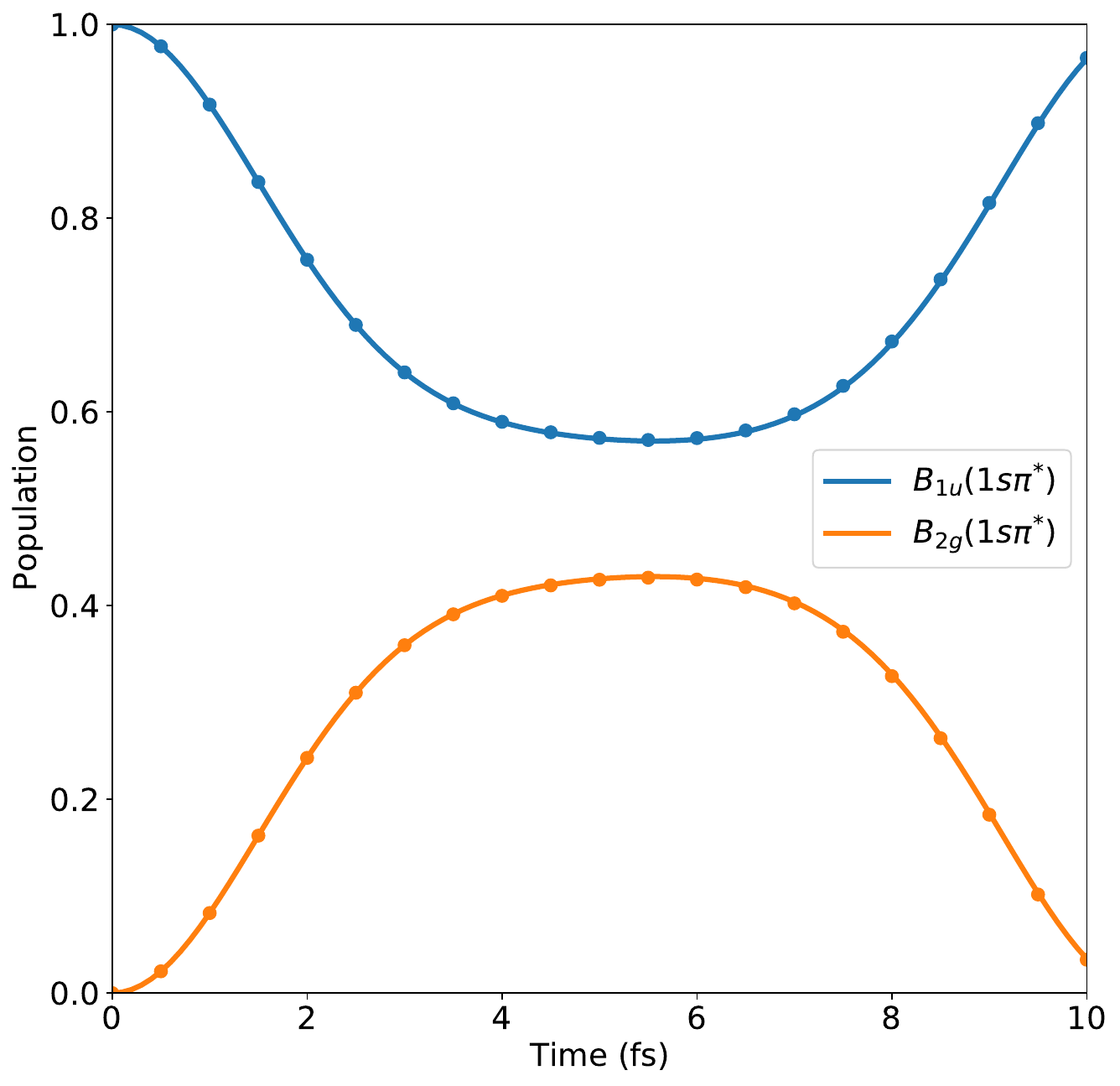}
    \caption{Diabatic state populations following excitation to the $B_{1u}(1s\pi^{*})$ state. Solid lines: results of a wave packet propagation including both effective modes $x$ and $y$. Dots: results from the inclusion of only the non-adiabatic coupling direction $y$, corresponding to asymmetric C-H stretching.}
    \label{fig:pop}
\end{figure}

We first note that the gradient difference mode $x$ is dominated by the $a_{g}$ symmetric C-C stretching mode, while the non-adiabatic coupling mode $y$ corresponds to almost entirely to the $b_{3u}$ asymmetric C-H stretching mode, in agreement with previous work\cite{gadea1991,koppel1997}. Next, we consider the diabatic state populations following excitation to the $B_{1u}(1s\pi^{*})$ state, which are shown in Figure~\ref{fig:pop}. Here, solid lines are the results of a two-mode calculation, including both $x$ and $y$, whilst the dots are the result of a one-mode calculation including $y$ only. The initially excited $B_{1u}(1s\pi^{*})$ state undergoes rapid depopulation due to: (i) strong coupling to the $B_{2g}(1s\pi^{*})$ state, and; (ii) the near-zero energy difference between the two states. By 5.5~fs, 44\% of the population has been transferred to the $B_{2g}(1s\pi^{*})$ state, where after population flows back into the initially excited $B_{1u}(1s\pi^{*})$ state. These dynamics occur within the Auger decay lifetime, which is estimated to be around 7~fs\cite{larkins1996}. Remarkably, the gradient difference mode $x$ has essentially zero effect on the population dynamics, a consequence of a small difference in the $B_{1u}(1s\pi^{*})$ and $B_{2g}(1s\pi^{*})$ potential gradients. That is, the two diabatic potentials remain quasi-degenerate along $x$. Thus, as far as the population contribution to the vibronic core-hole density, $\Delta\rho_{pop}^{(core)}(\mathbf{r},\boldsymbol{R},t)$, goes, this mode may be safely neglected. Furthermore, as the gradient difference is the principle mechanism for decoherence, this also means that the effect of the mode $x$ on the coherence contribution, $\Delta\rho_{coh}^{(core)}(\mathbf{r},\boldsymbol{R},t)$, will also be negligible. Thus, to an excellent approximation, the core-hole dynamics can be modeled using a simple one-mode model including only the non-adiabatic coupling direction $y$, corresponding to asymmetric C-H stretching.

To visualise the core-hole localisation dynamics, we consider what we shall term the ``core-hole asymmetry", $\delta(\boldsymbol{R},t)$, corresponding to the difference of the vibronic core-hole density projected onto the `left' and `right' hand sides of the molecule, which is encoded in the following quantity:

\begin{equation}
    \begin{aligned}
        \delta(\boldsymbol{R},t) &= \int d\mathbf{r} \Big[ \hat{P}_{1s_{r}}(\mathbf{r};\boldsymbol{R}) \Delta\rho^{(core)}(\mathbf{r},\boldsymbol{R},t)\\
        &- \hat{P}_{1s_{l}}(\mathbf{r};\boldsymbol{R}) \Delta\rho^{(core)}(\mathbf{r},\boldsymbol{R},t) \Big],
    \end{aligned}
\end{equation}

\noindent
where $\hat{P}_{1s_{l}}(\mathbf{r};\boldsymbol{R})$ and $\hat{P}_{1s_{r}}(\mathbf{r};\boldsymbol{R})$ denote the projectors onto the left/right carbon $1s$ atomic orbitals, which are localised on the left- and right-hand-side carbon atoms and denoted by $1s_{l}$ and $1s_{r}$ here and in the following. The core-hole asymmetry $\delta(\boldsymbol{R},t)$ gives the difference between the integral (over electronic coordinates) of the vibronic core-hole density projected onto the spaces spanned by the two different localised $1s$ orbitals. A negative value corresponds to a net accumulation of core-hole density on the left-hand-side carbon atom, and positive values to a net accumulation on the right-hand-side carbon atom. A value of zero corresponds to a completely delocalised vibronic core-hole density.

\begin{figure*}
    \centering
    \includegraphics[width=0.4\textwidth]{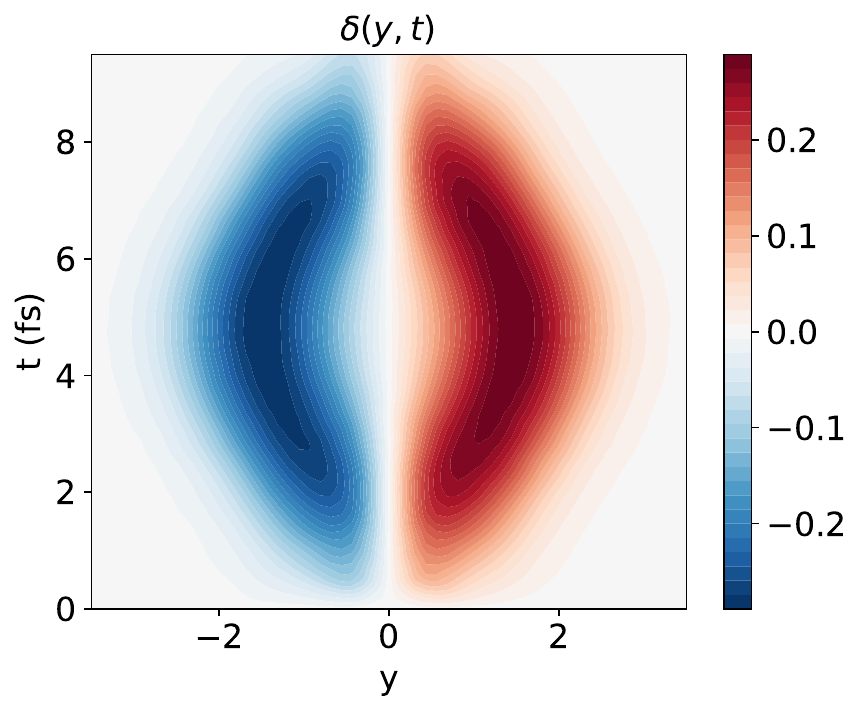}
    \includegraphics[width=0.4\textwidth]{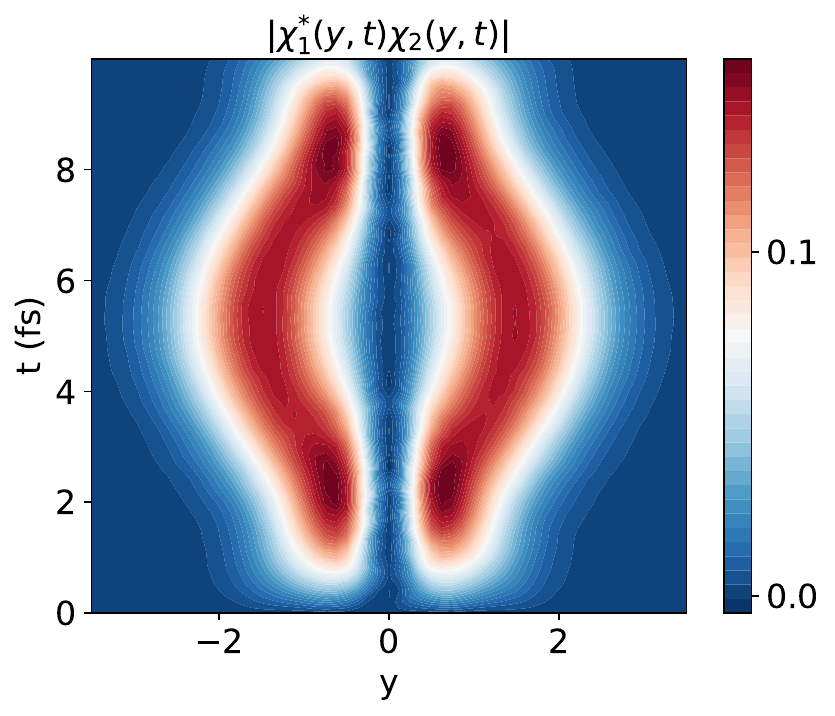}
    \caption{Left: the core-hole asymmetry $\delta(y,t)$ following excitation to the $B_{1u}(1s\pi^{*})$ state. Negative values correspond to a net accumulation of core-hole density on the left-hand-side carbon atom, and positive values to a net accumulation on the right-hand-side carbon atom. Right: absolute value of the product of the nuclear wave functions for the $B_{1u}(1s\pi^{*})$ and $B_{2g}(1s\pi^{*})$ states.}
    \label{fig:hole_dens}
\end{figure*}

The left-hand panel in Figure~\ref{fig:hole_dens} shows the core-hole asymmetry $\delta(y,t)$ computed following vertical excitation to the $B_{1u}(1s\pi^{*})$ state. At time $t=0$, the core-hole asymmetry is zero for all geometries, corresponding to a vibronic core-hole density $\Delta\rho^{(core)}(\mathbf{r},\boldsymbol{R},t)$ is that completely delocalised. This in turn is a result of only the $B_{1u}(1s\pi^{*})$ state being populated, and the coherence contribution $\Delta\rho^{(core)}_{coh}(\mathbf{r},\boldsymbol{R},t)$ being zero. By around 2~fs, the core-hole becomes partially localised, as shown by the development of small, but non-zero core-hole asymmetry values at displaced values of the coupling mode $y$. This is caused by the transfer of population to the $B_{2g}(1s\pi^{*})$ state and the coherence contribution $\Delta\rho_{coh}^{(core)}(\mathbf{r},\boldsymbol{R},t)$ to the core-hole density beginning to grow. However, a significant nuclear probability density still exists at the $D_{2h}$ ground state minimum energy geometry ($y=0$), at which the core-hole remains delocalised. By 5.5~fs, when the difference in the $B_{1u}(1s\pi^{*})$ and $B_{2g}(1s\pi^{*})$ state populations approaches a minimum, and the coherence contribution reaches its maximum, the localisation of the vibronic core-hole density reaches its maximum, as reflected in the large values of $\delta(y,t)$ at displaced values of the coupling mode $y$. These core-hole dynamics map directly onto the time-evolution of the nuclear wave function products $\chi_{1}^{*}(y,t)\chi_{2}(y,t)$, the absolute values of which are shown in the right-hand panel of Figure~\ref{fig:hole_dens}. We note that, due to symmetry, the value of $\langle \chi_{1} | \chi_{2} \rangle$ is zero for all time\cite{coherences_ours}, however, the local values of $\chi_{1}^{*}(y,t)\chi_{2}(y,t)$ at any particular value of $y$ maybe non-zero.

\begin{figure}
    \centering
    \includegraphics[width=0.4\textwidth]{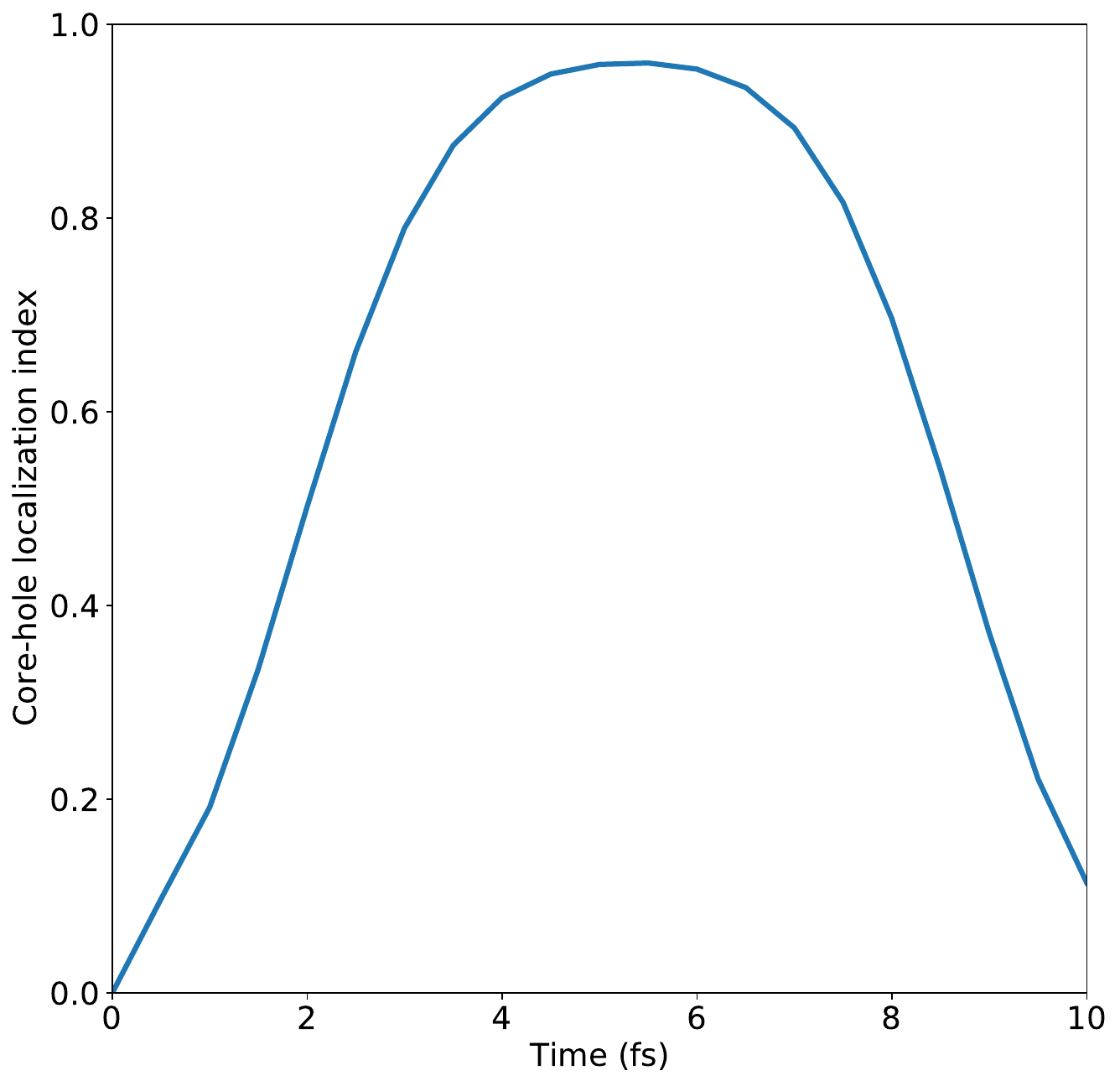}
    \caption{Core-hole localisation index $L(t)$ following excitation to the $B_{1u}(1s\pi^{*})$ state. A value of zero corresponds to complete core-hole delocalisation over the two carbon atoms, and a value of one to complete core-hole localisation on a single carbon atom.}
    \label{fig:loc_index}
\end{figure}

To quantify the degree of core-hole localisation, we introduce the core-hole localisation index $L(t)$, defined as

\begin{equation}
    L(t) = \int d\boldsymbol{R} \text{ abs}\left[ \delta(\boldsymbol{R},t) \right],
\end{equation}

\noindent
The localisation index $L(t)$ corresponds to the integral of the absolute value of the core-hole asymmetry, $\delta(\boldsymbol{R},t)$, over nuclear coordinates. For singly-core-excited states within the CVS approximation, this has limiting values of zero and one, corresponding to complete delocalisation and localisation of the core-hole density, respectively. The core-hole localisation index following excitation to the $B_{1u}(1s\pi^{*})$ state is shown in Figure~\ref{fig:loc_index}. At time $t=0$, the core-hole density is completely delocalised over the two carbon atoms, a result of only one of the two electronic states being populated and the population contribution, $\Delta\rho_{pop}^{(core)}(\mathbf{r},\boldsymbol{R},t)$, dominating. Rapidly, however, core-level charge transfer occurs, causing the core-hole to localise, which is encoded in the increasing value of the core-hole localisation index $L(t)$. This increase tracks the population dynamics, and the value of $L(t)$ reaches a maximum of 0.96 at 5.5~fs, corresponding to an almost entirely localised core hole. This is time at which the population is most equally spread over the $B_{1u}(1s\pi^{*})$ and $B_{2g}(1s\pi^{*})$ states. Importantly, these dynamics occur within the estimated core-hole lifetime of around 7~fs\cite{larkins1996}. Therein after, the core hole begins to delocalise again as population flows back into the initially excited $B_{1u}(1s\pi^{*})$ state.

To conclude, we have demonstrated that few-fs core-level charge transfer may occur following X-ray absorption, leading to ultrafast core-hole localisation. Using the example of ethylene excited to its $1s\pi^{*}$ manifold by an ultrafast X-ray pump pulse, we predict that the initially delocalised core-hole will undergo near-total localisation within around 5~fs, driven by core-level charge transfer. This is followed by core-hole delocalisation, all of which occurs within the Auger decay window. The mechanism for these dynamics results from a breakdown of the Born-Oppenheimer approximation, and is driven by high-frequency asymmetric C-H stretching. This runs contrary to conventional wisdom, which assumes that charge transfer will occur on longer timescales than charge migration since the former is necessarily driven by nuclear motion, while the latter involves the purely electronic dynamics that ensue following the preparation of a coherent superposition electronic states. It is interesting to note that, in both cases, the electronic dynamics are driven by the formation of coherences. In the cases of charge migration, these are created by the pump laser pulse, and nuclear dynamics can lead to their rapid destruction\cite{vacher2017}. In the case considered here, however, it is the nuclear motion that creates the coherences and drives the electronic dynamics. We anticipate that this phenomenon will be common to any molecule for which manifolds of nearly-degenerate core-excited or core-ionised states exist, corresponding to the excitation of two or more equivalent $1s$ core orbitals which, for organic molecules with multiple equivalent atoms is the rule rather than the exception. Moreover, as the core-level charge transfer process occurs within the Auger decay window, it is in principle observable using modern X-ray light sources\cite{huang2017, vendrell2024}.



%

\end{document}